\documentclass{svmult}

\usepackage{mathptmx}       
\usepackage{helvet}         
\usepackage{courier}        
\usepackage{type1cm}        
\usepackage{makeidx}         
\usepackage{graphicx}        
\usepackage{multicol}        
\usepackage[bottom]{footmisc}

\usepackage{amsmath}

\makeindex             

\begin{document}

\title*{Estimation of Discretized Motion of Pedestrians by the Decision-Making Model}
\titlerunning{Estimation of Discretized Motion by DM Model}
\author{Pavel Hrab\'ak and Ond\v rej Tich\'a\v cek and Vladim\'ira Se\v{c}k\'arov\'a}
\institute{P. Hrab\'ak, O. Tich\'a\v cek and V. Se\v{c}k\'arov\'a \at Institute of Information Theory and Automation, Czech Academy of Sciences, Pod Vodarenskou vezi 4, CZ-182 08, Prague 8, Czech Republic.  \email{hrabak@utia.cas.cz}
}
%
%
\maketitle

\abstract{The contribution gives a micro-structural insight into the pedestrian decision process during an egress situation. A method how to extract the decisions of pedestrians from the trajectories recorded during the experiments is introduced. The underlying Markov decision process is estimated using the finite mixture approximation. Furthermore, the results of this estimation can be used as an input to the optimization of a Markov decision process for one `clever' agent. This agent optimizes his strategy of motion with respect to different reward functions, minimizing the time spent in the room or minimizing the amount of inhaled CO.
}

\section{Introduction}
\label{sec:Introduction}

This study can be used as an auxiliary calibration tool for microscopic models of pedestrian flow with spatially discretized motion of agents, as e.g. floor-field model~\cite{KirSch2002PhysicaA} or optimal-steps model~\cite{SeiKoe2012PRE}. The results can be applied in the navigation robotic systems~\cite{SpaMatVeiTiaLimPed2015AAMA}. The introduced method builds upon the floor-field model. Thanks to the restriction to the discretized motion of pedestrian we are able to express the local decisions of pedestrian in the terms of Markov decision process~\cite{Puterman1994}.

For the analysis of the real data, we use the experimental data from a passing-through experiment~\cite{BukHraKrb2014Procedia}. In this experiment, pedestrians were instructed to pass through a simple room equipped by one entrance with controlled inflow and one exit of the width 60~cm. Since we are mainly interested in the pedestrian interaction, we used the data from the rear camera covering the space of 2.5~m in front of the exit and short part of the corridor behind the exit.

Throughout the article, we use the notation related to Markov decision processes (MDP) adopted from the book~\cite{Puterman1994}. The main task of the contribution is to express the basic entries of the MDP theory in the scope of pedestrian flow dynamics. This is necessary to use the optimization technique described in~\cite[Chapter 4]{Puterman1994}.

\section{Basic Concept}

Let us describe the MDP in general. The considered decision process (DP) is characterized by a sequence $(s_1,a_1,s_2,a_2,\dots,s_{T-1},a_{T-1},s_T)$ of states $s_t\in S$ and performed actions $a_t\in A$. Here $T$ plays the role of a finite time horizon used for the optimization. At time $t$ an agent, who is making the decision, observes the system to be in state $s_t$ and based on this observation performs an action $a_t$ with conditional probability $p_t(a_t\mid s_t)$. The system reacts to the action stochastically and the state changes to $s_{t+1}$ with conditional probability $p_t(s_{t+1}\mid s_t, a_t)$. This probability can be understood as the agent's image of the environment behaviour. The Markov property is hidden in the fact that both, the decision part $p_t(a_t\mid s_t)$ and the environmental model $p_t(s_{t+1}\mid s_t, a_t)$, depend only on the situation at time $t$. Then the probability of a sequence $(s_1,a_1,s_2,a_2,\dots,s_{T-1},a_{T-1},s_T)$ is given as
\begin{equation}
	\Pr\left(s_1,a_1,s_2,a_2,\dots,s_{T-1},a_{T-1},s_{T}\right)=p(s_1)\prod_{t=1}^{T-1} p_t(a_t\mid s_t)p_t(s_{t+1}\mid s_t,a_t)
\end{equation}

This concept can be easily applied to the floor-field model~\cite{KirSch2002PhysicaA} (For more details about the model we refer the reader to the book~\cite{SchChoNis2010}). The floor-field model is a particle hopping model defined on a rectangular lattice $L$ representing the discretized layout of the simulated facility. Particles are hopping between cells stochastically according to the hopping probabilities, which are influenced by the static floor field $S$. Usually $S(y)=\operatorname{dist}(y,\vec{E})$ refers to the distance of the cell $y$ to the exit $\vec{E}$ in defined metric $\operatorname{dist}$. Let the state of the system at time $t$ be denoted by $\tau_t\in\{0,1\}^{L}$, where $\tau(x)=1$ refers to an occupied cell and $\tau(x)=0$ to an empty cell. Let further $n_t=\sum_x\tau_t(x)$ be the number of agents in the lattice at time $t$.

In each algorithm step $t\to t+1$, every agent $i\in \{1,\dots,n_t\}$ chooses his future position $y_{i,t}$ given he is sitting in $x_{i,t}$ with probability
\begin{equation}
\label{eq:pks}
	p(y_{i,t}\mid x_{i,t}, \tau_t)\propto\exp\{-S(y_{i,t})\}\mathbf{1}_{\{\operatorname{dist}(x_{i,t},y_{i,t})\leq1\}}
\end{equation}
according to floor-field model. The model of the environment is then a consequence of the choices of future positions of all agents, i.e., the dynamics is driven by the environment model
\begin{equation}
\label{eq:beh}
	p\left(\tau_{t+1}\mid\tau_t, y_{i,t}, i\in\{0,\dots,n_t\}\right) = F(p(y_{i,t}\mid x_{i,t}, \tau_t), i\in\{0,\dots,n_t\})\,,
\end{equation}
where $F$ is a function that reflects conflicting situation where two agents choose the same target cell\footnote{Without the conflicting situations the function is just a product of the entries.}.

\section{Estimating ${p_t(a_t\mid s_t)}$}
\label{sec:est}

This section focuses on the probability probabilistic decision ${p_t(a_t\mid s_t)}$ from the data recorded during evacuation experiment~\cite{BukHraKrb2014Procedia}.

\begin{figure}[htb]
	\centering
	\begin{minipage}{0.33\textwidth}
		\centering		
	\includegraphics[height=3.3cm]{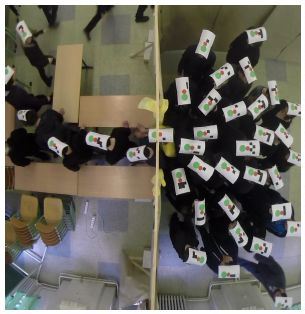}
	\end{minipage}%
	\begin{minipage}{0.31\textwidth}
		\centering		
		\includegraphics[clip, trim={0.9cm 1cm 0.8cm 0.7cm},height=3.3cm]{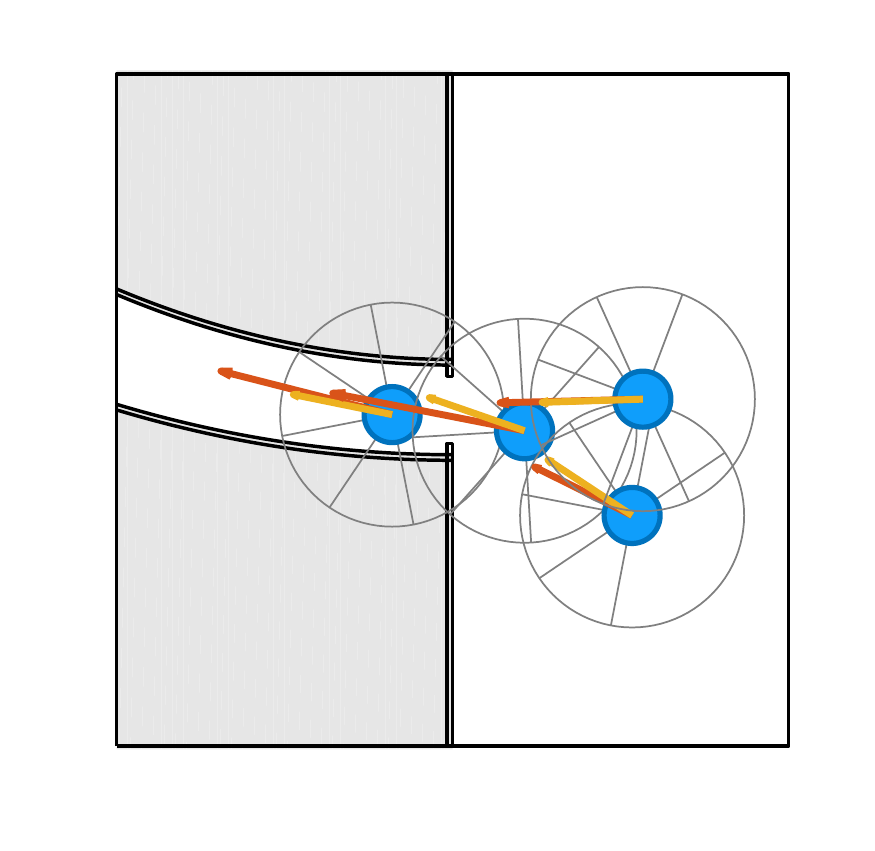}
	\end{minipage}%
	\begin{minipage}{0.32\textwidth}
		\centering		
		\includegraphics[height=3.3cm]{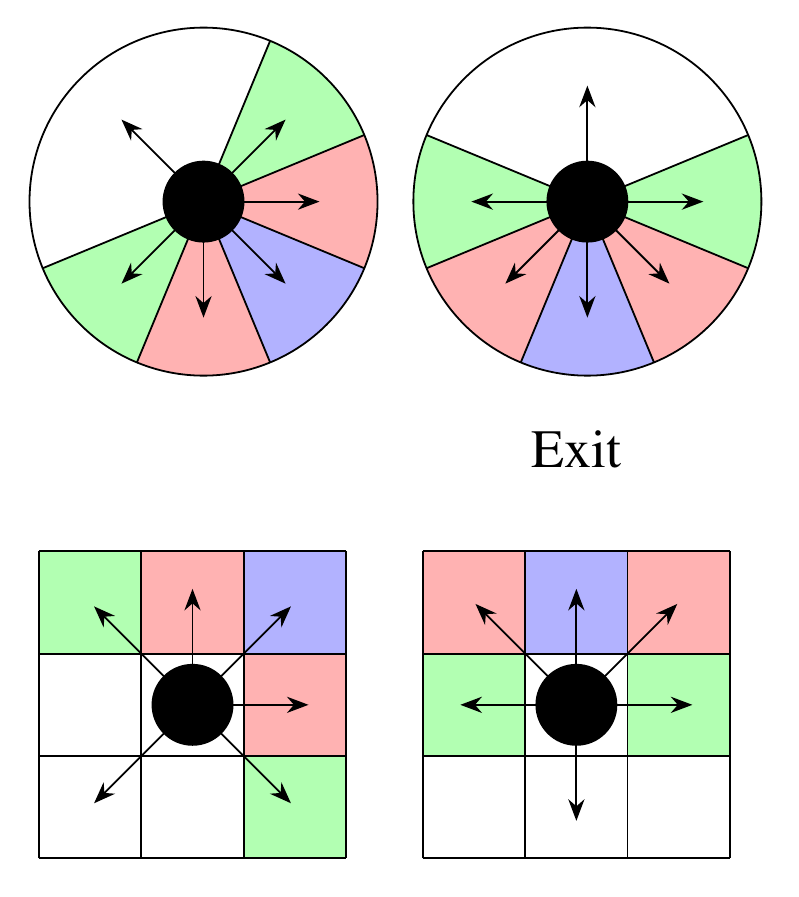}
	\end{minipage}
\caption{Transformation of trajectory record to actions. Trajectories are extracted from the video records (left, taken from~\cite{BukHraKrb2014Procedia}), transformed to motion within sectors (midle) with respect to the direction towards exit, and interpreted as motion in the lattice (right).}
\label{fig:exp}
\end{figure}

Let us assume that the pedestrians act similarly to the floor-field particles, i.e., all pedestrians are following the same decision strategy, which does not change in time and space. Furthermore, we assume that pedestrians react only on their immediate neighbourhood reflecting the direction towards the exit, but not their absolute position. Therefore, the state $s_t$ in MDP can be associated with the state of the immediate neighbourhood. Contrary to the floor-field model we consider the neighbourhood to be oriented with respect to the direction towards the exit. The actions are associated with direction angle a pedestrian can choose, see Figure~\ref{fig:exp}.

The experimental data for trajectory analyses have been provided by our colleague Marek Buk\'a\v{c}ek (Czech Technical University). The data are in the form of paths records $\left(\vec{x}_i(t), t\in[t_i^\mathrm{in},t_i^\mathrm{out}]\right)$, where $t_i^\mathrm{in}$ and $t_i^\mathrm{out}$ is the time of the first and the last appearance of the pedestrian $i$ on the screen respectively. $\vec{x}_i(t)$ is the position of the pedestrian on the screen at time $t$. To match the discrete nature of the decision-making process, the motion of pedestrians has been discretized in time with the discretization step $\Delta t=1$~s. The vector of motion at time $t$ is then $\Delta\vec{x}_i(t)=\vec{x}_i(t+\Delta t)-\vec{x}_i(t)$ and the direction of motion $\vartheta_i(t)$ is an angle given by
\begin{equation}
	\cos\vartheta_i(t)=\frac{[\vec{E}-\vec{x}_i(t)]\cdot\Delta\vec{x}_i(t)}{\|\vec{E}-\vec{x}_i(t)\| \cdot \|\Delta\vec{x}_i(t)\| }
\end{equation}

This angle is then associated to the action $a\in A$ according to the Table~\ref{tab:act}. The set of actions is $A=\{\otimes, \leftarrow, \nwarrow, \swarrow, \uparrow, \downarrow, \rightarrow\}$, e.g., an angle $\vartheta_i(t)=20^\circ$ corresponds to the forward motion $\leftarrow$, while $\vartheta_i(t)=30^\circ$ corresponds to the left-forward motion $\swarrow$. Every motion performed with velocity ${v}_{i}(t)=\|\Delta\vec{x}_i(t)\|/\Delta t$ less then 0.5~m/s has been considered as standing ($\otimes$). 

\begin{table}
\caption{Action set in detail. Direction angle towards the exit is 0. Colours refer to Figure~\ref{fig:exp}.}
\label{tab:act}
\centering
	\begin{tabular}{p{1.3cm}ccccccc}
		\hline\noalign{\smallskip}
			\textbf{Action:} & $\otimes$ & $\leftarrow$ & $\nwarrow$ & $\swarrow$ & $\uparrow$ & $\downarrow$ & $\rightarrow$  \\
		\noalign{\smallskip}\svhline\noalign{\smallskip}
			\textbf{Angle:} &$\emptyset$ & $\left(-\frac{\pi}{8},\frac{\pi}{8}\right)$ & $\left(-\frac{\pi}{8},-\frac{3\pi}{8}\right)$ & $\left(\frac{3\pi}{8},\frac{\pi}{8}\right)$& $\left(-\frac{3\pi}{8},-\frac{5\pi}{8}\right)$ & $\left(\frac{5\pi}{8},\frac{3\pi}{8}\right)$ & $\left(-\pi,-\frac{5\pi}{8}\right)\cup\left(\frac{5\pi}{8},\pi\right)$ \\
			\noalign{\smallskip}
		\textbf{Color:}& black&blue&red&red&green&green&white \\
		\noalign{\smallskip}\hline\noalign{\smallskip}
	\end{tabular}
\end{table}

\begin{figure}[htb]
	
	\begin{minipage}{0.29\textwidth}
		\centering
		{\scriptsize no filter}
	\end{minipage}%
	\hspace{0.04\textwidth}
	\begin{minipage}{0.29\textwidth}
		\centering
		{\scriptsize filter velocity $ < $ 0.5 m/s}
	\end{minipage}
	
	\begin{minipage}{0.31\textwidth}
		\centering		
		\includegraphics[clip, trim={1.2cm 0.25cm 0 0.5cm}, width=\textwidth]{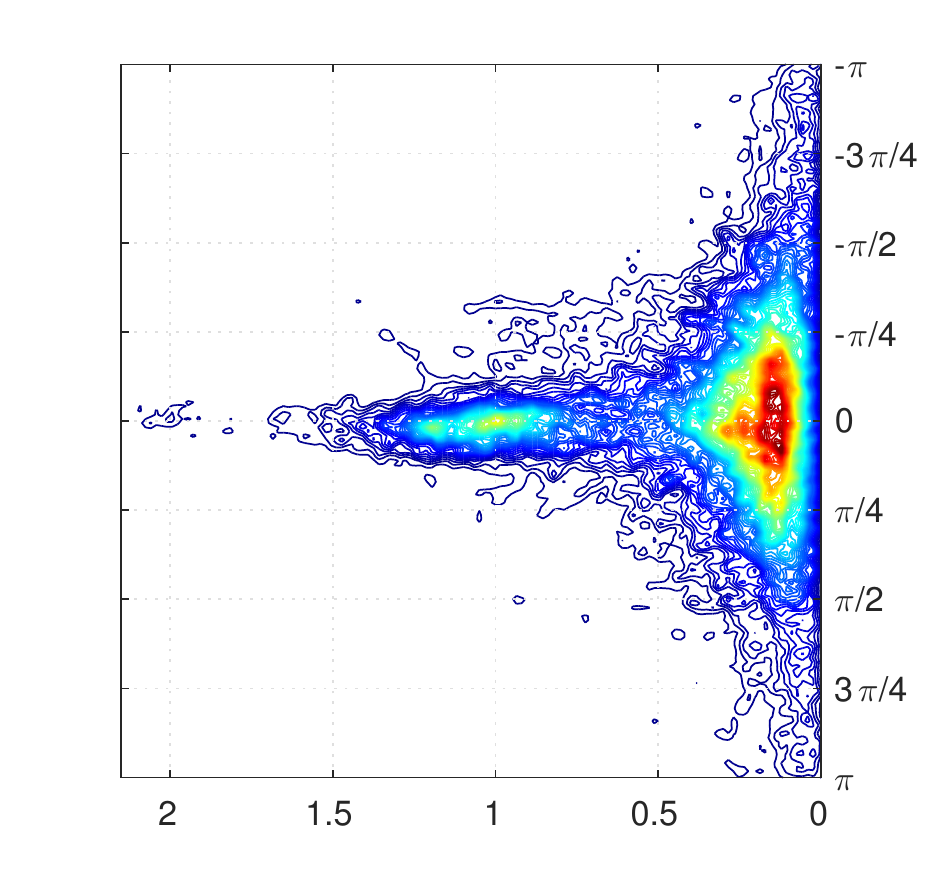}
	\end{minipage}%
		\rotatebox[origin=c]{90}{\scriptsize direction angle}	
	\begin{minipage}{0.31\textwidth}
		\centering		
		\includegraphics[clip, trim={1.2cm 0.25cm 0 0.5cm}, width=\textwidth]{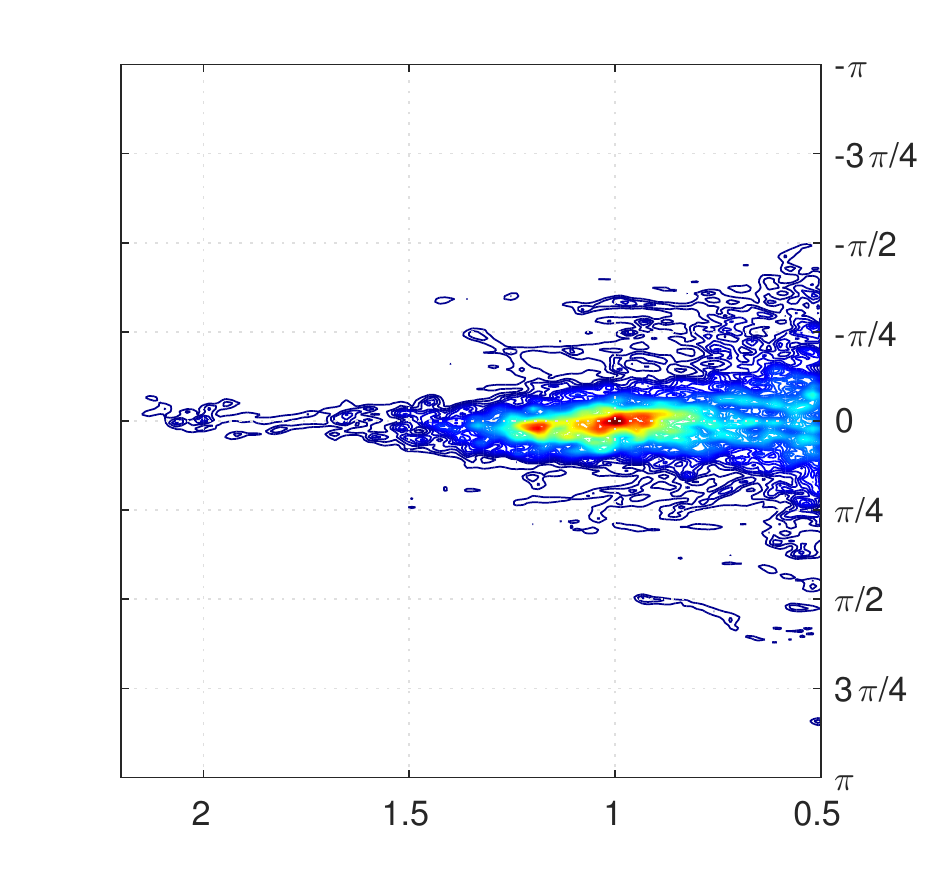}
	\end{minipage}%
		\centering	
		\rotatebox[origin=c]{90}{\scriptsize  direction angle}	
	\begin{minipage}{0.32\textwidth}
		\centering		
		\includegraphics[clip, trim={0.1cm 0.5cm 0.5cm 0.5cm}, width=\textwidth]{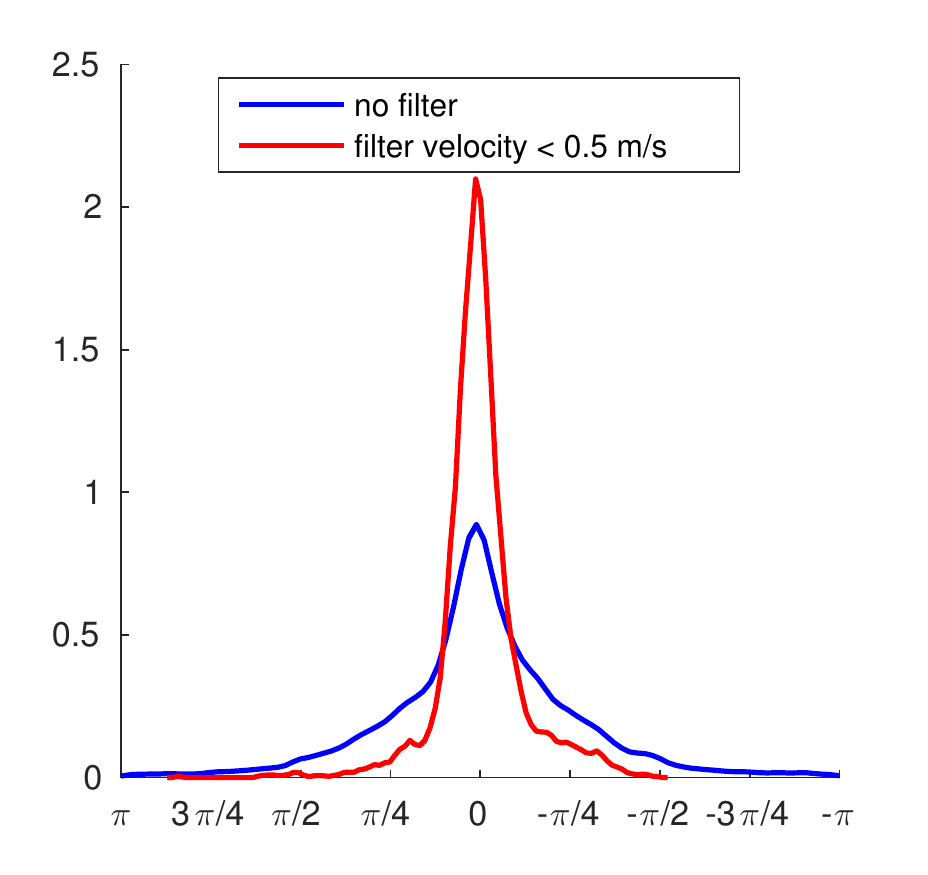}
	\end{minipage}

	\begin{minipage}{0.29\textwidth}
		\centering
		{\scriptsize movement length}
	\end{minipage}%
	\hspace{0.04\textwidth}
	\begin{minipage}{0.29\textwidth}
		\centering
		{\scriptsize movement length}
	\end{minipage}%
	\hspace{0.07\textwidth}	
	\begin{minipage}{0.28\textwidth}
		\centering
		{\scriptsize direction angle}
	\end{minipage}			
	
\caption{The frequency of occurrence of the motion length in given direction angle. The graphs are oriented similarly to the snapshot from the experiment in Figure~\ref{fig:exp}. Left: without the filter on standing. Middle: after filtering out the motion with velocity less then 0.5~m/s. Right: Gaussian-kernel estimation of the distribution of chosen direction.}
\label{fig:act}       
\end{figure}

In Figure~\ref{fig:act} the frequency of chosen direction with respect to the direction angle is plotted. Two main clusters for forward stepping  and standing  can be distinguished, the latter dominates. The data are aggregated over all pedestrians and all records for each. From this graph we can conclude that the majority of pedestrians preferred standing in line and moving forward centimetre by centimetre rather then trying to push through the crowd or overrunning it.

For the estimation of the decision process we associate the state $s_t$ in the decision-process $p(a_t\mid s_t)$ with the occupancy of the immediate neighbourhood. By the neighbourhood $N(\vec{x})$ of a position $\vec{x}$ we understand a circle around $\vec{x}$ with the radius 0.75~m (maximal step size) divided into 6 sectors $\{\leftarrow, \nwarrow, \swarrow, \uparrow, \downarrow, \rightarrow\}$ defined in previous section. The state $s_{i,t}$ for the decision $p(a_{i,t}\mid s_{i,t})$ of the agent $i$ is then a vector from $\{0,1\}^6$, where $s_{i,t}(y)=0$ for empty sector ${y}$ and $s_{i,t}(y)=1$ for occupied sector. Here $y\in\{\leftarrow, \nwarrow, \swarrow, \uparrow, \downarrow, \rightarrow\}$. The sector is considered occupied if it contains at least one position vector of another agent or if it is covered by a wall by at least 40~\%. Since the data are aggregated over all pedestrians, the index $i$ will be further omitted.

Most natural way how to estimate the decision process $p(a_t\mid s_t)$ is to compare the frequency of chosen directions (actions) given the occupancy of the neighbourhood. However, this method fails in the case of the trajectory data from considered experiment, since most of the combinations $(a_t,s_t)$ appear very rarely. For this reason we applied the approximation of the decision by finite mixture model with forgetting~\cite{Kar:14a}. The idea consists in approximation of the complex decision process $p(a_t\mid s_t)$ by the convex combination of marginal decision processes $\Theta(a_t|s_t(y))$, i.e.,
\begin{equation}
\label{eq:pat}
	p(a_t\mid s_t)\approx p(a_t \mid \Theta,s_t)=\sum_{y\in N(\vec{x}_t)}\alpha_y\Theta(a_t|s_t(y))\,,
\end{equation}
where $\alpha_y$ is the coefficient of influence of the state of $y$ to the decision; $\sum_{y}\alpha_y =1$. For more details see~\cite{HraTic:15}.

The resulting values of the mixture model are given in Table~\ref{tab:res}. The following phenomena can be observed analysing the values in the table. The occupancy of a neighbouring sector almost always contributes with the highest value to the decision ``to stand'' ($ \otimes $). However, a free neighbouring sector does not always tend to imply motion, see the forward sector ($ \leftarrow $). The most diverse influence of the empty and occupied states show the ``slightly right'' ($ \nwarrow $) and ``slightly left'' ($ \swarrow $) sectors. The explanation for this may be a zipper-like effect of agents passing through a narrow exit and corridor.

The table also implies, that the occupancy of right and left sectors ($ \uparrow, \downarrow $) does not play a significant role in agent's decision as it does not restrain him from moving in desired (forward) direction. Finally, although in principle, the occupancy of the back sector ($ \rightarrow $) should not affect the agent's decision in his desire to go straight, this sector is mostly occupied if the agent is in a high-density situation (eg. a jam) and therefore it's occupancy reflects the agent's (in)ability to move at all.

\setlength{\tabcolsep}{4.75pt}
\begin{table}
\caption{Influence of regressors $ y $ to decision $ a $ for two states of sector occupancy ($ \circ $ symbolizes empty sector, $ \bullet $ occupied). $ \widehat{\alpha} $ is the weight of the regressor.}
\label{tab:res}
	\centering
	\begin{tabular}{c|c c|c c|c c|c c|c c|c c}
		\hline
		$ y $ & \multicolumn{2}{c|}{$ \leftarrow $}  & \multicolumn{2}{c|}{$ \nwarrow $}  & \multicolumn{2}{c|}{$ \swarrow $} & \multicolumn{2}{c|}{$ \uparrow $} & \multicolumn{2}{c|}{$ \downarrow $} & \multicolumn{2}{c}{$ 
		\rightarrow $} \\
		
		$ \widehat{\alpha} $ & \multicolumn{2}{c|}{    \textbf{\small 0.9212}} & \multicolumn{2}{c|}{    \textbf{\small 0.0749}}  & \multicolumn{2}{c|}{    \textbf{\small 0.0014}} & \multicolumn{2}{c|}{ 0.0002} & \multicolumn{2}{c|}{ 
		   0.0002} & \multicolumn{2}{c}{    \textbf{\small 0.0021}} \\
		\svhline
		$ a $ & \multicolumn{1}{c}{$ \circ $} & \multicolumn{1}{c|}{$ \bullet $}  & \multicolumn{1}{c}{$ \circ $} & \multicolumn{1}{c|}{$ \bullet $}  & \multicolumn{1}{c}{$ \circ $} & \multicolumn{1}{c|}{$ \bullet $}  & \multicolumn
		{1}{c}{$ \circ $} & \multicolumn{1}{c|}{$ \bullet $}  & \multicolumn{1}{c}{$ \circ $} & \multicolumn{1}{c|}{$ \bullet $}  & \multicolumn{1}{c}{$ \circ $} & \multicolumn{1}{c}{$ \bullet $} \\
		$ \otimes $            &  \bf \small 0.76   &  \bf \small 0.94	& 0.03    & \bf \small 0.97 & 0.05  &  \bf \small  0.52 & 0.11    & 0.12  & 0.12   & 0.11   & 0.07   &  \bf \small 0.69 \\
		$ \leftarrow $         & \bf \small 0.22   & 0.06  & \bf \small 0.63  & 0.01   &  \bf \small 0.73  &  0.12  & 0.39    & 0.23 & 0.35   & 0.24   & \bf \small  0.68  & 0.10 \\
		$ \nwarrow $           & 0.01   & 0.01    & 0.02    & 0.01   & 0.05  &  0.06  & 0.09    & 0.11  & 0.10   & 0.11   & 0.06   & 0.04 \\
		$ \swarrow $           & 0.01   & 0.01 	  & \bf \small 0.27  & 0.00    &  0.04  & 0.08    & 0.09  & 0.11   & 0.10   & 0.11   & 0.06   & 0.04 \\
		$ \uparrow $           & 0.00    & 0.00   & 0.02   & 0.00   & 0.04  &  0.06  & 0.08    & 0.11  & 0.09   & 0.11   & 0.04   & 0.04 \\
		$ \downarrow $         & 0.00    & 0.00   & 0.02   & 0.00   & 0.03  &  0.05  & 0.08    & 0.11  & 0.08   & 0.11   & 0.04   & 0.03 \\
		$ \rightarrow $        & 0.00    & 0.00   & 0.02   & 0.00   & 0.06  &  0.10  & 0.15    & 0.20  & 0.16   & 0.22   & 0.06   & 0.06 \\
		\hline
	\end{tabular}
\end{table}

\section{Optimizing ${p_t(a_t\mid s_t)}$}

This chapter offers an alternative view on the application of MDP to pedestrian flow modelling. Let the result of previous section be used as the behavioural frame of the majority of pedestrians, which determines the environmental model $p_t(s_{t+1}\mid s_t,a_t)$ described by equation~\eqref{eq:beh}. The aim of this section is to equip one of the pedestrian agent by an optimal decision strategy how to move among the pedestrians following the majority behaviour.

Let us in the following, for simplicity, return back to the Floor-field basis of the simulation. Consider that there is one ``clever'' particle among the ordinary undistinguishable floor-field particles behaving according to~\eqref{eq:pks} or~\eqref{eq:pat}. By the optimal strategy of the clever particle we understand the sequence $(d_1,d_2,\dots,d_{T-1})$, where $d_t(s_t)=q$ is the distribution on the set of actions $A$ playing the role of the decision process, i.e., $p(a_t\mid s_t)=q(a_t)$. The strategy is optimized with respect to given reward function $R(s_1,a_1,\dots,s_T)$, which can be used to model different preferences of the clever particle.

\begin{figure}[htb]
\centering
	\includegraphics[height=3.7cm]{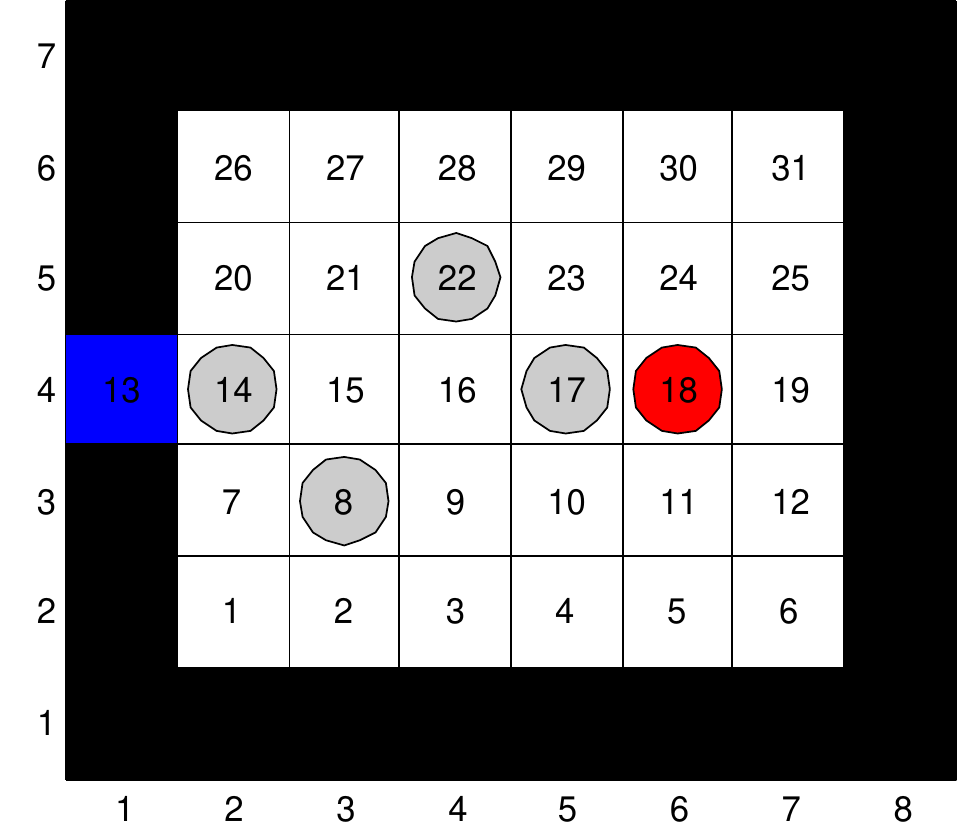}
	\hspace{1cm}
	\includegraphics[height=3.7cm]{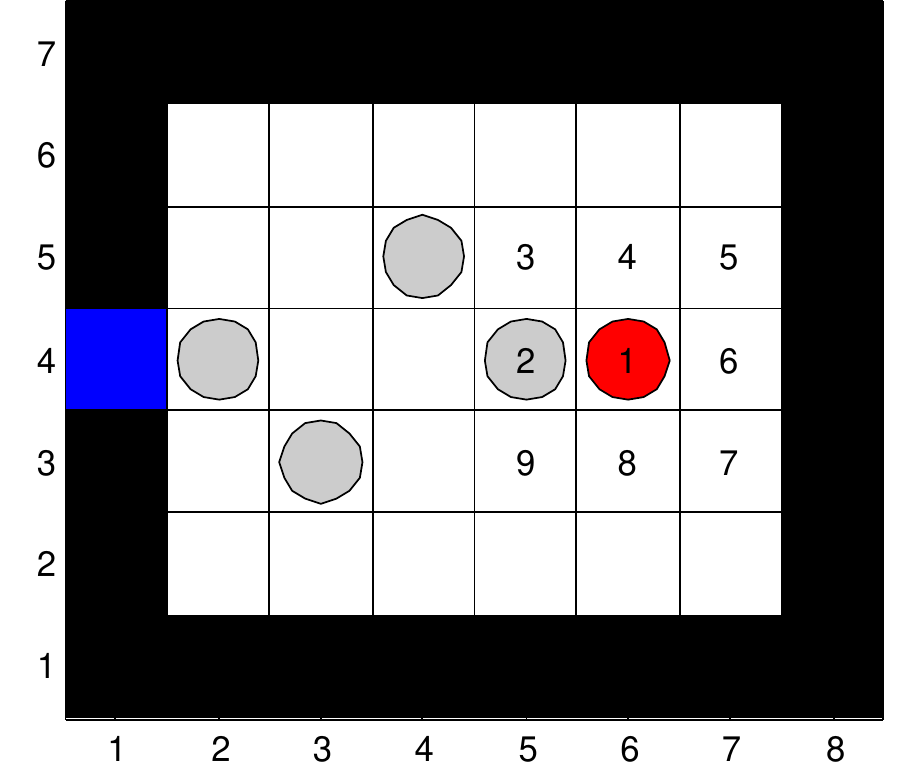}
\caption{An example of the state of the lattice with one clever agent (red) and 4 floor-field particles (gray). Left: the state is $s_t=(x_t,z_t)$, where  $x_t = 18$, $z_t=\{8,14,17,22\}$. Right: the numbering of directions corresponding to actions $a_t\in\{1,\dots,9\}$.}
\label{fig:state}       
\end{figure}

Contrarily to Section~\ref{sec:est} we consider in the following the position and the orientation to be absolute, i.e., there are not preferred positions on the lattice regarding the distance or orientation towards the lattice. The optimizing algorithm is supposed to find the shortest path given the maximal reward $R$ itself. The state of the system is expressed by the position of the clever agent $x_t$ and by the set of positions of all floor-field particles $z_t$, where $|z_t|=n_t-1$ (compare to previous section, where only the neighbouring pedestrians play role). Let the positions be numbered by natural numbers as shown by an example in Figure~\ref{fig:state}.

The actions an agent can choose are related to the 8 neighbouring sites and a possibility to stay in current position. Therefore the action set can be chosen as $A=\{1,2,\dots,9\}$, where the directions are numbered as depicted in Figure~\ref{fig:state}, i.e., $a_t=1$ means to choose as next target site the current position $x_t=18$; $a_t=2$ corresponds to the target site $x_t$`+'$(-1,0)=17$, etc. Here we note that, similarly to the floor-field, the chosen target site can be entered by a near floor-field particle. The choice of the target site is therefore influenced by the probability that other particles can change their positions.

For the purposes of this contribution we have chosen a simple updating scheme in which the clever agent moves after all other particles performed their actions. Then the environmental model $p_t(s_{t+1}\mid s_t,a_t)=p(x_{t+1},z_{t+1}\mid x_t,z_t,a_t)$ decomposes into a stochastic part $p(z_{t+1}\mid x_t,z_t)$ and a deterministic part $p(x_{t+1}\mid x_t,a_t,z_{t+1})$ as
\begin{equation}
	p(x_{t+1},z_{t+1}\mid x_t,z_t,a_t)=p(z_{t+1}\mid x_t,z_t)p(x_{t+1}\mid x_t,a_t,z_{t+1})\,.
\end{equation}	
In our case, the transition probability~\eqref{eq:pat} can be simplified to the form
\begin{equation}
	p(z_{t+1}\mid x_t,z_t)\propto\exp\{-U(z_{t+1})\}\,,\qquad U(z_{t+1})=\sum_{j=1}^{|z_{t+1}|}\mathrm{dist}(z_{t+1}(j),\vec{E})
\end{equation}
for all states $z_{t+1}$ reachable from $z_t$ by the motion of floor-field particles by one site. The introduced potential $U$ supports the states in which particles are closer to the exit and therefore suppresses the random motion away from the exit.

The above mentioned concept fits the finite time optimization of the MDP strategies using the backward induction algorithm described in~\cite[Section 4.5]{Puterman1994}. The final step is to define properly the reward function $R$. The reward function is defined as a cumulation of local rewards $r_t(s_t,a_t)$ and the final reward $v_T(s_T)$, i.e.,
\begin{equation}
	R=\sum_{t=1}^{T-1}r_t(s_t,a_t)+v_T(s_T)\,,\qquad r_t(s_t,a_t)=\sum_{s_{t+1}}p(s_{t+1}\mid s_t,a_t)r_t(s_{t+1},s_t,a_t)\,.
\end{equation}

The final reward is the same for all agents preferences taking into account the distance to the exit multiplied by a factor of 2, i.e., $v_T(s_T)=-2\operatorname{dist}(x_T,\vec{E})$. The local reward then reflects the agent's preferences. We introduce two main approaches: minimizing the time spent in the room and minimizing the amount of inhaled  carbon monoxide (CO) related to the aim to minimize number of lost conflicts.

The reward function minimizing the time simply subtracts one reward unit for each step an agent spends outside the exit, i.e.,
\begin{equation}
	r_t(s_{t+1},s_t,a_t)=\begin{cases}
				0 & x=\vec{E}\,,\\
				-1 & x\neq \vec{E}\,.
			\end{cases}	
\end{equation}
The reward function minimizing the amount of inhaled CO takes into account the possibility that the agent can choose a site which becomes occupied by another particle. Such choice can be interpreted as running to another pedestrian, which causes a significant loss of energy with no improvement of the distance to the exit. Such situation costs 2 reward units, while standing only one half. Therefore
\begin{equation}
	r_t(s_{t+1},s_t,a_t)=\begin{cases}
				0 & x_t=\vec{E}\,,\\
				-1/2 & a_t=1\,,\\
				-1 &  a_t\neq1\,,\,x_{t+1}\notin z_{t+1}\,,\\
				-2 &  a_t\neq1\,,\,x_{t+1}\in z_{t+1}\,.
			\end{cases}	
\end{equation}

\section{Conclusions and Future Plans}

The main goal of this paper was to introduce a concept of Markov decision process (MDP) to the pedestrian flow simulation. Two aspects have been studied by means of this concept: the estimation of pedestrian behaviour within crowded area and the optimization of the decision with respect to given pedestrian preferences. Both approaches are motivated by the cellular floor-field model used for simulation of pedestrian evacuation.

The estimation of pedestrian behaviour have been analysed from experimental trajectories. By means of the space discretization and finite mixture approximation we have been able to extract the pedestrians decision in relation to the occupation of his immediate neighbourhood. The analysis showed that the main influence to the decision has the occupation of the area in the forward direction towards the exit. Further, most of the decisions pedestrians performed was to move forward or stay at the position. The over-running of the crowd was rather a rare event.

The results of the experiment analyses can be then used as the input to the optimization task of one `clever' agent among usual floor-field particles. We have introduced a technique of expressing the pedestrian evacuation model in terms of the MDP. Furthermore, two different reward functions have been introduced to simulate different preferences of the clever agent: to minimize the time spent in the room and to minimize the amount of inhaled carbon oxygen, i.e. minimizing the number of conflicts. In the future we plan to test the combination of those two strategies in order to prove  that optimal is the combination  of the two above mentioned strategies.
\begin{acknowledgement}
	This work was supported by the Czech Science Foundation under the grant GA13-13502S. We want to thank our colleague Marek Buk\'a\v{c}ek from Czech Technical University for the provided experimental data.
\end{acknowledgement}

\bibliographystyle{spmpsci}
\bibliography{TGF15_PH+OT+VS_bibtex}

\begin{thebibliography}{1}
\providecommand{\url}[1]{{#1}}
\providecommand{\urlprefix}{URL }
\expandafter\ifx\csname urlstyle\endcsname\relax
  \providecommand{\doi}[1]{DOI~\discretionary{}{}{}#1}\else
  \providecommand{\doi}{DOI~\discretionary{}{}{}\begingroup
  \urlstyle{rm}\Url}\fi

\bibitem{BukHraKrb2014Procedia}
Buk\'a\v{c}ek, M., Hrab\'ak, P., Krb\'alek, M.: Experimental study of phase
  transition in pedestrian flow.
\newblock In: W.~Daamen, D.C. Duives, S.P. Hoogendoorn (eds.) Pedestrian and
  Evacuation Dynamics 2014, \emph{Transportation Research Procedia}, vol.~2,
  pp. 105--113. Elsevier Science B.V. (2014)

\bibitem{HraTic:15}
Hrab{\'{a}}k, P., Tich{\'{a}}{\v{c}}ek, O.: Prediction of pedestrian decisions
  during the egress situation: Application of recursive estimation of
  high-order {M}arkov chains using the approximation by finite mixtures.
\newblock Tech. Rep. 2346, \'{U}TIA AV \v{C}R, POBox 18, 182 08 Prague 8, Czech
  Republic (2015)

\bibitem{Kar:14a}
K\'arn\'y, M.: Recursive estimation of high-order markov chains: Approximation
  by finite mixtures.
\newblock Information Sciences \textbf{326}, 188--201 (2016)

\bibitem{KirSch2002PhysicaA}
Kirchner, A., Schadschneider, A.: Simulation of evacuation processes using a
  bionics-inspired cellular automaton model for pedestrian dynamics.
\newblock Physica A: Statistical Mechanics and its Applications
  \textbf{312}(1–2), 260 -- 276 (2002)

\bibitem{Puterman1994}
Puterman, M.L.: Markov Decision Processes: Discrete Stochastic Dynamic
  Programming, 1st edn.
\newblock John Wiley \& Sons, Inc., New York, NY, USA (1994)

\bibitem{SchChoNis2010}
Schadschneider, A., Chowdhury, D., Nishinari, K.: Stochastic Transport in
  Complex Systems: From Molecules to Vehicles.
\newblock Elsevier Science B. V., Amsterdam (2010)

\bibitem{SeiKoe2012PRE}
Seitz, M.J., K\"oster, G.: Natural discretization of pedestrian movement in
  continuous space.
\newblock Physical Review E \textbf{86}, 046,108 (2012)

\bibitem{SpaMatVeiTiaLimPed2015AAMA}
Spaan, M., Veiga, T., Lima, P.: Decision-theoretic planning under uncertainty
  with information rewards for active cooperative perception.
\newblock Autonomous Agents and Multi-Agent Systems \textbf{29}(6), 1157--1185
  (2015)

\end{thebibliography}

\end{document}